\documentclass[preprint]{aastex}
\pdfoutput=1 
\slugcomment{Submitted to A\&A on Feb 6, 2015} 
\usepackage{rotating}
\usepackage{natbib}
\usepackage{lscape}
\usepackage{color}
\tighten

\newcommand{\mjup}{\mbox{$M_{Jup}$}}

\newcommand{\mic}{\mbox{$\mu$m}}
\newcommand{\app}{\mbox{$\sim$}}

\newcommand{\ro}{\mbox{$\rho$}}

\newcommand{\dg}{\mbox{$^\circ$}}

\newcommand{\prim}{\mbox{HD~135344~B}}

\newcommand{\eg}{e.g.}
\newcommand{\ie}{i.e.}

\makeatletter
\newcommand{\Rmnum}[1]{\expandafter\@slowromancap\romannumeral #1@}
\makeatother

\begin{document}


\title {Improving Signal to Noise in the Direct Imaging of Exoplanets
  and Circumstellar Disks}

\author {Zahed Wahhaj\altaffilmark{1}, 
  Lucas A. Cieza\altaffilmark{2},
  Dimitri Mawet\altaffilmark{1},
  Bin Yang\altaffilmark{1},
  Hector Canovas\altaffilmark{3},
  Jos De Boer\altaffilmark{1},
  Simon Casassus\altaffilmark{4},
  Fran\c cois M\'enard\altaffilmark{4,8}, 
  Matthias R. Schreiber\altaffilmark{3},
  Michael C. Liu\altaffilmark{5},  
  Beth A. Biller\altaffilmark{6},
  Eric L. Nielsen\altaffilmark{5},
  and Thomas L. Hayward\altaffilmark{7}
}

\altaffiltext{1} {European Southern Observatory, Alonso de Cordova 3107,  Vitacura, Casilla 19001, Santiago, Chile}
\altaffiltext{2} {Facultad de Ingenieria, Universidad Diego Portales. Av. Ejercito 441, Santiago, Chile}
\altaffiltext{3} {Departamento de Fsica y Astronoma, Universidad Valparaiso, Avenida Gran Bretana 1111, Valparaiso, Chile}
\altaffiltext{4} {Departamento de Astronomía, Universidad de Chile, Casilla 36-D, Santiago, Chile}
\altaffiltext{5} {Institute for Astronomy, University of Hawaii, 2680 Woodlawn Drive, Honolulu, HI 96822}
\altaffiltext{6} {Institute for Astronomy, University of Edinburgh, Blackford Hill, Edinburgh EH9 3HJ, UK}
\altaffiltext{7} {Gemini Observatory, Southern Operations Center, c/o AURA, Casilla 603, La Serena, Chile}
\altaffiltext{8} {UMI-FCA, CNRS/INSU, France (UMI 3386)}


\begin{abstract}
We present a new algorithm designed to improve the signal to 
noise ratio (SNR) of point and extended source detections in direct imaging data. 
The novel part of our method is that it finds the linear combination of the science images that best
match counterpart images with signal removed from suspected source regions.
The algorithm, based on the Locally Optimized Combination of Images (LOCI)
method, is called {\it Matched} LOCI or MLOCI. 
\textcolor{black}{
We show using data
obtained with the Gemini Planet Imager (GPI) and Near-Infrared Coronagraphic Imager
(NICI) that the new algorithm can improve the SNR of point source detections
by 30--400\% over past methods. We also find no increase in  false
detections rates. No prior knowledge of candidate companion locations is
required to use MLOCI.    
While non-blind applications may yield linear combinations of science images
which seem to increase the SNR of true sources by a
factor $>$2, they can also yield false detections at high rates.}
This is a potential pitfall when trying to confirm marginal detections or
to re-detect point sources found in previous epochs.  
Our findings are relevant to any method where the coefficients
of the linear combination are considered tunable,
e.g. LOCI and Principal Component Analysis (PCA). Thus we recommend
that false detection rates be analyzed when using these techniques. 
\end{abstract}

\footnotetext[0]{Based on observations obtained at the Gemini
  Observatory, which is operated by the Association of Universities for
 Research in Astronomy, Inc., under a cooperative agreement with the
  NSF on behalf of the Gemini partnership: the National Science
  Foundation (United States), the Science and Technology Facilities
  Council (United Kingdom), the National Research Council (Canada),
  CONICYT (Chile), the Australian Research Council (Australia),
  Minist\'{e}rio da Ci\^{e}ncia e Tecnologia (Brazil) and Ministerio de
  Ciencia, Tecnolog\'{i}a e Innovaci\'{o}n Productiva (Argentina).}


\section {INTRODUCTION}

\footnotetext[1]{http://exoplanet.eu/catalog/} 

While more than 1700 planets have been detected by transit and radial
velocity techniques, only about a dozen planetary-mass objects ($\leq$
13 \mjup ) have been directly imaged around stars to date\footnotemark .
However, this number is expected to increase significantly 
\citep[\eg][]{2011PASP..123..692M} since first light has just been obtained 
with the new extreme adaptive optics instruments, Gemini Planet
Imager\citep[GPI;][]{2008SPIE.7015E..31M} and SPHERE \citep{2010ASPC..430..231B}. A severe challenge in AO imaging arises from quasi-static
speckles, which can emulate astronomical point sources for hour long
timescales \citep[\eg][]{2005JRASC..99..130M}.
Besides high-order AO corrections and coronagraphy, the most powerful advances to
overcome this sensitivity barrier are (1) angular differential imaging \citep[ADI;][]{2004Sci...305.1442L, 2005JRASC..99..130M} 
which decouples the sky rotation of the planet from the speckles, and (2)
spectral difference imaging \citep[SDI;][]{1999PASP..111..587R,2007ApJS..173..143B} and spectral deconvolution 
\citep{2002ApJ...578..543S,2007MNRAS.378.1229T} which decouple  the star and 
planet spectrally, taking advantage of the fact that speckles move out
radially with wavelength and that planet and stellar spectra differ. 
To obtain the maximum SNR for true companions, the speckle structure, which is correlated over time and wavelength,
must be subtracted. Thus, some science images must take negative weight in the linear
combination to create the final reduced image.
Tremendous effort  has
been dedicated to finding algorithms that optimize detections by varying the
weights of science images. The most prominent of these have been the LOCI algorithm \citep{2007ApJ...660..770L},
and the PCA based algorithms, KLIP
\citep{2012ApJ...755L..28S} and PynPoint \citep{2012MNRAS.427..948A},
which have focussed on minimizing speckle residuals.
Subsequent improvements have concentrated on optimizing SNR of point
source recovery, \eg\ damped LOCI, ALOCI and TLOCI \citep{2012ApJS..199....6P,2013ApJ...776...15C,2014arXiv1407.2555M}.

Our own experiments with data from the Gemini NICI Planet-Finding
Campaign \citep{2010SPIE.7736E..53L} based on blind
recovery of simulated companions showed no
measurable improvement in achieved contrast using LOCI compared to
conventional ADI \citep{2013ApJ...779...80W}. We
found that the main problem with LOCI was that reductions in noise
around detections were also accompanied by reductions in the signal.
Here, we present a LOCI variant that 
significantly improves SNR. We demonstrate the effectiveness of the algorithm using NICI
Campaign datasets and the early science GPI dataset on  \prim , a transition disk with strong evidence for ongoing planet formation.  
The science results from the GPI observations will be published in Menard et al.\ (2015). 
 
\section{MATCHED LOCI}


The ADI reduction method is based on the principle that when the image
rotator is turned off for an altitude-azimuth telescope, at the
Cassegrain focus, the sky rotates with respect to the detector,
while the quasi-static speckle pattern caused by the telescope and
instrument optics do not. At the Nasmyth focus, one achieves the same
effect in an observing mode called pupil-tracking. Astrophysical sources are thus separated from
the speckle pattern, which is isolated by median combining a series of
speckle-aligned images to create a reference PSF. The reference PSF is
then subtracted from each science image. The difference images 
are then de-rotated to align the sky and stacked again to create the
final reduced image. The details of our own implementation of regular ADI 
can be found in \citet{2013ApJ...779...80W}. The LOCI algorithm
differs in the PSF subtraction step where it subtracts the best
linear combination of speckle-aligned images 
such that the RMS in some region of the Field of
View (FoV) is minimized.

\begin{figure}
    \includegraphics[width=18cm]{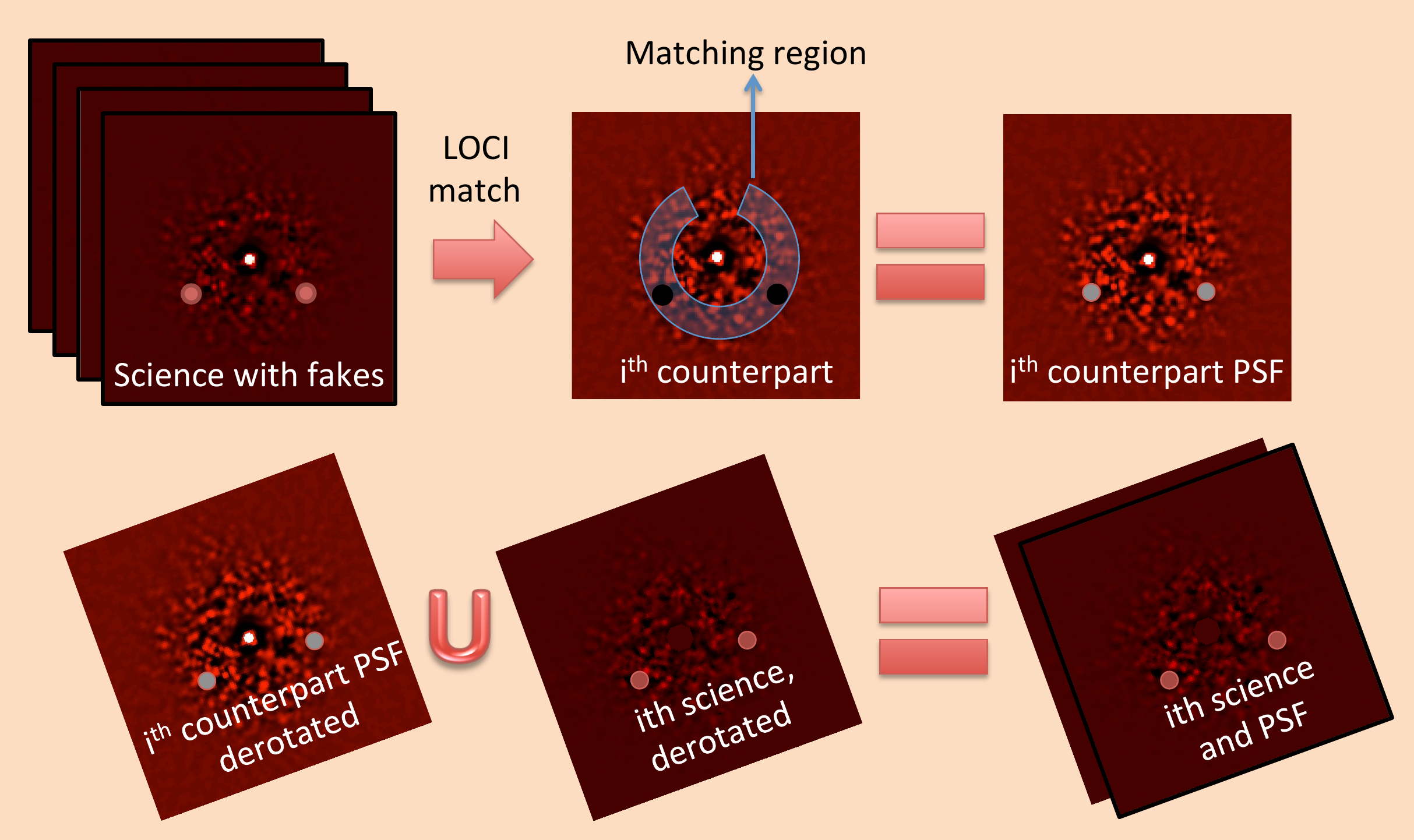}
    \caption{\textcolor{black}{The central steps of the MLOCI algortihm (see \S2). 
      Step 4, top row: For each science image a counterpart PSF is
      created. The PSF is such that its subtraction from the science
      preserves signal at 2 locations. Step 5, bottom row: Each
      science image and its counterpart PSF are de-rotated to align
      North Up.}}   
  \label{fig:schema}
\end{figure}

\begin{figure}[ht]
  \begin{center}
    \includegraphics[width=18cm]{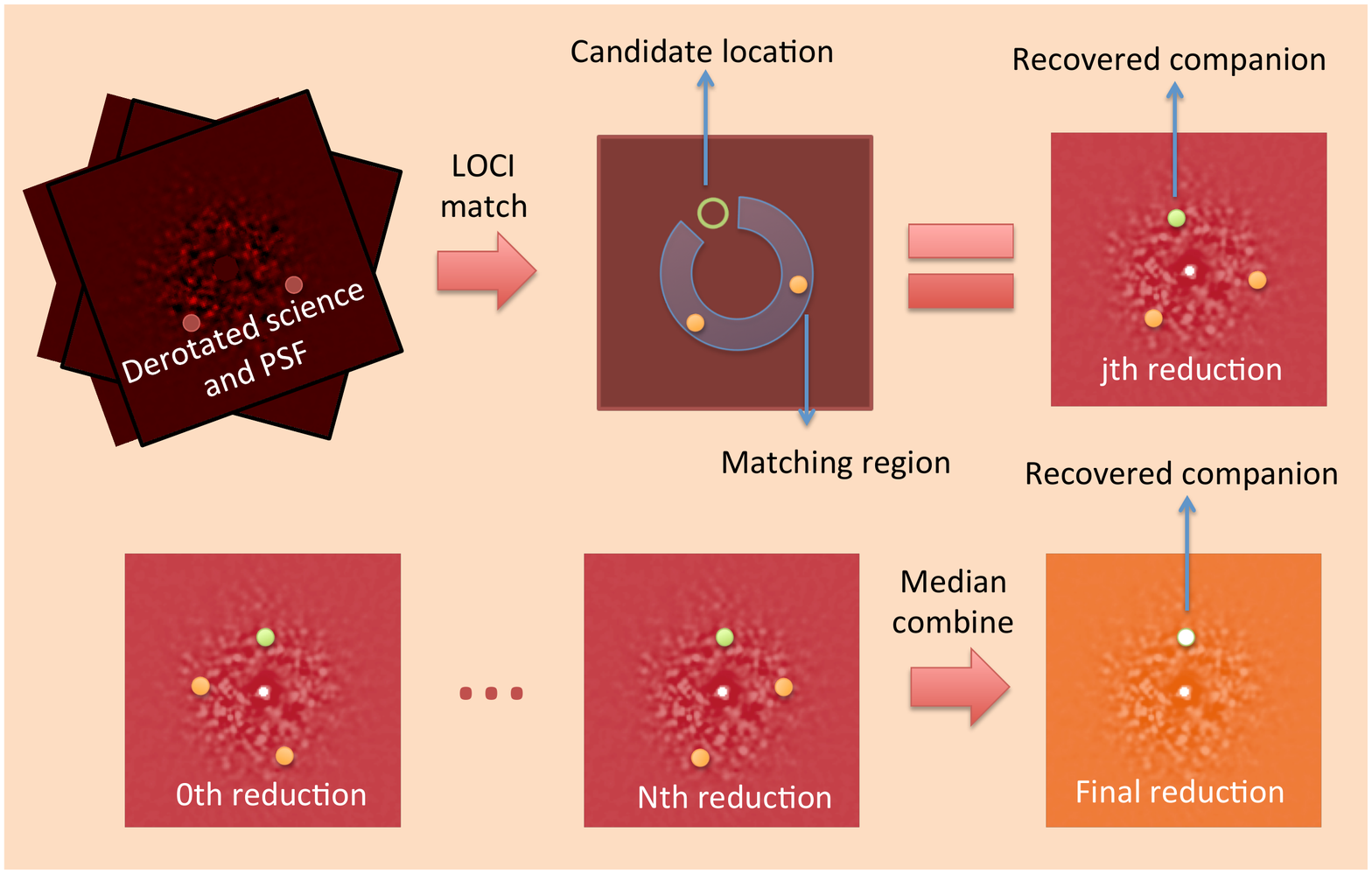}
    \end{center}
    \caption{\textcolor{black}{The final steps of the MLOCI algorithm (see \S2).
      Step 7, top row: Instead of the usual difference image combination
      in ADI, all science and PSF images are combined to recover
      injected sources in reference sector. Step 9, bottom row:
      Several such reductions created through step 7, are combined to
      create the final reduction. The final reduction preserves point
      source signal, generally.}}   
  \label{fig:schema2}
\end{figure}

\textcolor{black}{
Below we list the steps of our new algorithm which we call MLOCI or
Matched LOCI:
\begin{enumerate}
\item 
A basic ADI reduction is initially applied to identify all candidate point sources
with SNR$>$2.  
\item 
For each candidate, we define an annulus with inner and outer radii
bounding the candidate. In this annulus, 
a sector subtending 40 degrees is defined with the candidate
azimuthally centered and radially one PSF FWHM farther out than the
inner edge. The rest of the annulus, subtending 320 degrees, is defined as the reference sector.
\item     
For each science image, we create a counterpart where a point source is
subtracted from two (or three) sky locations in the reference sector
(see Figure~\ref{fig:schema}). The point sources which we subtract are
simulated to undergo ADI self-subtraction, but for 30\% less rotation as would occur in the ADI
data set. However, the sources are not allowed to shrink azimuthally
to widths $<$0.7$\times$FWHM.  The radial width of the sectors, the
number of point sources subtracted and their intensity, we have found by
experiments which we describe in \S4.
\item 
Next, LOCI is used to linearly combine the original science images to find the best
match to each counterpart image to make counterpart PSFs. The science
images should be speckle-aligned at this stage. When there are several
spectral channels, the images need to be spectrally deconvolved as
described in \S3. The matching is done only over the reference sector, not the target sector which
includes the candidate location. 
\item 
The science images and their counterpart PSFs are then derotated to
align North Up.
\item 
An image with two (or three) positive sources at the sky locations of
the previous negative sources (see Step 3) is made. The background of this image is
zero \ie\ it is noiseless (Figure~\ref{fig:schema2}, top row, middle) .
\item  
Then LOCI is used to match the positive source image by linearly combining the derotated science images
and counterpart PSFs (see Figure~\ref{fig:schema2}). When there are hundreds of images (too time-comsuming for LOCI), we process subsets
of the images and median combine the results.
\item  
The sky locations of the injected sources 
are changed to positions not over-lapping with the previous ones.
The steps above are repeated to obtain 5--10 different reductions.
Since the injected sources are only needed to compute the coefficients of
the linear combination, for the final combinations
we use the original images without the source injections. 
Thus, these 5--10 reductions only contain real astronomical sources
and not the injected ones.
\item  
The reductions above are then median combined to create the final
reduced image for a particular annular region.  
\item  
For X candidates, X annular reductions are made, the widths of which
are at least as large as those of the matching annuli. 
The outer radius of these annuli are extended up to the matching annulus
(inner radius) of the candidate next in radial separation.
When the annular reductions overlap due to densely packed candidates,
the overlapping pixels are median combined. 
\end{enumerate}
}

\textcolor{black}{
It is important to remember 
that MLOCI only uses injected point sources to solve for the
linear coefficients for image combination. Point sources are not
injected at the locations of candidate companions but in the reference
sectors which exclude the candidates.
}

\textcolor{black}{
We also present here an implementation of MLOCI for extended sources,
which one could apply if there is any evidence for extended emission
in a basic ADI reduction.
However, in this implementation the occurence of false positives
becomes a concern. Therefore, an MLOCI detection of a data set 
devoid of signal should be performed to estimate the likelihood of
false positives. Moreover, as a test, one can attempt to recover injected 
extended sources in empty data sets using MLOCI. 
This method is most promising when a library of PSFs from a different star are available
as references, thus allowing one to avoid self-subtraction of the
extended source. However, such detailed analyses are defered to a
later publication. 
}

\textcolor{black}{
Below we list the steps of MLOCI for extended source detection:
\begin{enumerate}
\item 
A preliminary reduction (\eg\ basic ADI) is done first to identify all
regions likely containing signal (\eg\  SNR $>$ 2). These regions are defined as the target,
while the regions which are likely devoid of signal are defined as
background (\eg\ SNR $<$ 1). These regions together make the reference region and will
be matched by MLOCI. Other non-reference regions in the image will not be
matched. Although, this step depends on prior estimates of the emmisive
and non-emmisive regions around the target, we show that the final reduction
is not very likely to yield false disk detections (see \S4.3). 
\item     
For each science image, we create a counterpart where the local 
RMS is subtracted from the target sky regions. 
\item 
Then LOCI is used to linearly combine the original science images to find the best
match to each counterpart image to make counterpart PSFs. As in the
point-source reduction,
the science images need to be speckle-aligned at this stage. The matching
is done only over the reference region.
\item 
The science images and their counterpart PSFs are then derotated to
align North Up.
\item 
The preliminary reduction is then set to zero at the background sky
regions to make the final counterpart. 
\item  
Then LOCI is used to match the final counterpart by linearly combining the derotated science images
and counterpart PSFs.  When there hundred of images (too many for LOCI), we process subsets
of the images and median combine the results.
\end{enumerate}
}

\section{OBSERVATIONS AND DATA REDUCTION}
We observed \prim\ in the $J$-band on April 21st, 2014 UT as part of GPI Early Science
(Program ID = GS-2014A-SV-402). \prim\ is a prime
example of a transition disk suspected to harbor a substellar
companion (a brown dwarf or a recently formed giant planet). 
Submillimeter images show a dust-depleted cavity with a radius of $\app$40
AU in a nearly face-on disk, while polarimetric differential imaging
in the near-IR reveals two remarkable spiral arms extending from 
$\app$28 AU (inside the submillimeter cavity) to $\app$130 AU \citep{2013A&A...560A.105G}.

We obtained images from the integral field spectrograph (IFS)
in 37 spectral channels from $\lambda$=1.1--1.35~\mic\ using GPI's
unprecedented adaptive optics capabilities. 
The star was placed behind a coronagraphic mask of radius 92~mas and thus a
properly detected point source has projected separation, \ro$\geq$112~mas (since $J$-band
resolution is \app 40~mas~\footnote{http://www.gemini.edu/sciops/instruments/gpi/instrument-performance?q=node/12172}). 
The IFS has a plate scale of 14.3~mas/pixel and a 2.8$''\times$2.8$''$ field of view (FoV). The
telescope rotator was turned off allowing the sky to rotate on the
detector through 35\dg , so that the speckle pattern produced by the telescope and
instrument optics would be decoupled from real astronomical sources.
This ensures a total azimuthal sky motion of 1.7$\times$FWHM at 112~mas (and 2$\times$FWHM at
130~mas). We obtained 39 sixty-second exposures. A total of 50 minutes
of telescope time excluding acquisition was used. A speckle at 112~mas should
move outward by only $(\frac{1.35}{1.1}-1)\times$112~mas=25.5~mas ($<$1$\times$FHWM) across images
obtained over $\lambda=$1.1--1.35~\mic\ and thus we do not expect much
contrast gain from spectral deconvolution at this separation. 

We used the GPI pipeline \citep{2010SPIE.7735E..31M} to produce wavelength-calibrated
spectral cubes with bad-pixel removal, de-striping, flat-fielding,
non-linearity, and persistence corrections.
The centroid of each of the 37$\times$39$=$1443 images are found by averaging the coordinates of
the peaks of the four GPI satellites 
spots~\footnote{http://www.gemini.edu/sciops/instruments/gpi/public-data/public-data-readme}. 
The ADI datasets from each spectral channel are reduced separately
using the pipeline described in \citet{2013ApJ...779...80W}. 
However, we do not spatially filter the science images. Also,
the reference PSFs are not translated to optimize the fit to the science images
before subtraction, but only scaled in intensity to minimize the RMS
between \ro $=$70--210~mas. 
The 37 ADI reductions from each of the spectral channels are then median combined. 

\textcolor{black}{
For the MLOCI reduction, we create counterpart PSFs for each of 
the 1443 science images. Before combining science images to create the
counterpart PSFs, they have to be spectrally de-convolved. 
That is, to match an image in the $i$th channel, the science images are
magnified by a factor $\lambda_i/\lambda_j$, where $\lambda_i$ and $\lambda_j$ are
the central wavelengths of the $i$th and $j$th channels,
respectively. The science images and their PSF counterparts images are then
derotated to align North Up.    
We then use LOCI again to match the adjusted preliminary reduction, by
linearly combining the science images and the counterpart PSFs. 
We have two counterpart PSFs for each science image: one made from
images in the same spectral channel, and the other made from images
with the same orientation on detector. 
In the last step, the 1443$\times$3 science and counterpart
PSF images can be combined using LOCI to match the final counterpart
(See \S2).}  However, this is often too many images for
LOCI to combine. So instead, we sequentially combine all images
(science and two counterpart PSFs) in the $j$th channel and $j$th
orientation with (37+39)$\times$3=228 images in each combination.  
See Figure~\ref{fig:gpi_reducs} for the reduction of the GPI data.

We also demonstrate the effectiveness of the MLOCI algorithm using 10
methane band\\ ($CH_4S,~\lambda=$1.578~\mic ) datasets from the Gemini NICI
Planet-Finding Campaign \citep[2008--2012;][]{2010SPIE.7736E..53L,
  2013ApJ...773..179W, 2013ApJ...776....4N, 2013ApJ...777..160B}. 
The filter was designed to help detect methane which is usually found
in the atmospheres of cool substellar companions with surface temperatures $<$1400~K.
In this paper, the simultaneously exposed images in filter $CH_4L$ are not 
used for spectral differencing  as we want to keep the 
analysis simple. However, since we use a single narrow-band (4\% width) 
filter, we reach shallower contrasts than in the Campaign.
The Campaign targets used were HD~107146, HD~25457, HD~53143
, GJ~388, HD~31295, HIP~25486, HD~92945, HD~21997, HD~110058 and UY~Pic, all observed between January and March of 2009.
NICI was installed at the 8.1~m Gemini South Telescope 
and produced AO corrected images over a 18.4$''\times$18.4$''$ FoV detector.
The NICI plate scale was 17.96~mas/pixel. The Campaign observations used a translucent coronagraphic
mask \citep[central attenuation factor=358; ][]{2011ApJ...729..139W} with a half-power radius of
0.32$''$. Median contrast achieved at \ro =0$''$.5  was 12.6~mag \citep{2013ApJ...779...80W}. 
We already demonstrated in \citet{2013ApJ...779...80W} that the LOCI
pipeline does not perform better than regular ADI for NICI data.

We compare MLOCI reductions of the NICI datasets to both PCA
reductions using the KLIP algorithm \citep[see][]{2012ApJ...755L..28S}
and regular ADI reductions using the implementation described above.
The PCA algorithm creates an orthogonal basis of images which are then
linearly combined to emulate the supplied PSF images. Moreover, the 
basis are ordered such that the first basis or mode is the best
general match to the majority of the images, while each subsequent mode represents
perturbations of decreasing importance. Thus fewer modes can be used
in the hopes of reducing noise without removing too much signal, and
improve over LOCI. The PSF subtraction can also
be optimized for sub-regions of a science image.
For our PCA reductions, we obtained the best results (best recovery
SNR  for injected companions) for number of KLIP modes = 8, 
and PSF subtraction optimized for annuli of 
width 20 pixels. The inner radius of the innermost annulus was set to 25 pixels.

\section{RESULTS}
\subsection{Recovery of Simulated Companions}

\textcolor{black}{
To tune MLOCI and measure its optimum performance, we insert simulated companions
into  our 10 NICI datasets and recover them using MLOCI. We place the companions at 0$''$.5-1$''$.0 separations from the
primary (where sensitivity is speckle-limited) with contrasts just below the 5$\sigma$ detection limit for
regular ADI reduction. 
For each data set between 0$''$.5-1$''$.0, we have roughly 800 independent sky locations at
which to do this test, given that the FWHM of the NICI $H$-band PSF is
$\sim$55~mas. As described in \S2, injected test companions are bounded
by target sectors, while two or three point sources are inserted
into the corresponding reference sectors (annlus $-$ target sector) to aid MLOCI in preserving point source
signal. Our aim is to compare the SNR of simulated companions as recovered by
MLOCI, to the SNR as in recoveries by ADI and PCA reductions.
We measure MLOCI's false positive rate by repeating the reductions without simulated companions at the test locations.
}

First, we find the best reduction parameters for MLOCI given different
amount of sky rotations in an ADI dataset.
We multiply the true position angles (PA) of the FoV by a negative number to
misalign the images as done in \citet{2013ApJ...779...80W} so that any real 
astronomical sources will be mostly removed in a median combination of
the images. To simulate different amounts of sky rotation, we choose 5
values of this negative number so that the total sky motions at the
separations of interest, which we
call $\alpha$, are 0.35, 1.25, 2.15, 3.1 or  4.0 $\times$FWHM. We had to use
companions 2 and 0.5 mag brighter for $\alpha$ = 0.35 and 1.25 $\times$FWHM respectively,
because of reduced sensitivity in these cases due to large self-subtraction factors.
We also vary  $W$, the width of the reference sectors, over 4, 8, 16, 32 and 64 pixels and $P$ the pull
over 1,2,3,4 and 5. Here, $P$ is the brightness relative to the local RMS, of the point
sources that are subtracted from the reference sectors (see
\S~2). Reductions of the UY Pic dataset by varying the parameters over
this grid of values show that optimum settings are near $W=$16, $P=$3, while two simulated companions
in the reference sector to assist MLOCI are adequate. 

\begin{figure}[ht]
  \centerline{
      \includegraphics[width=12cm]{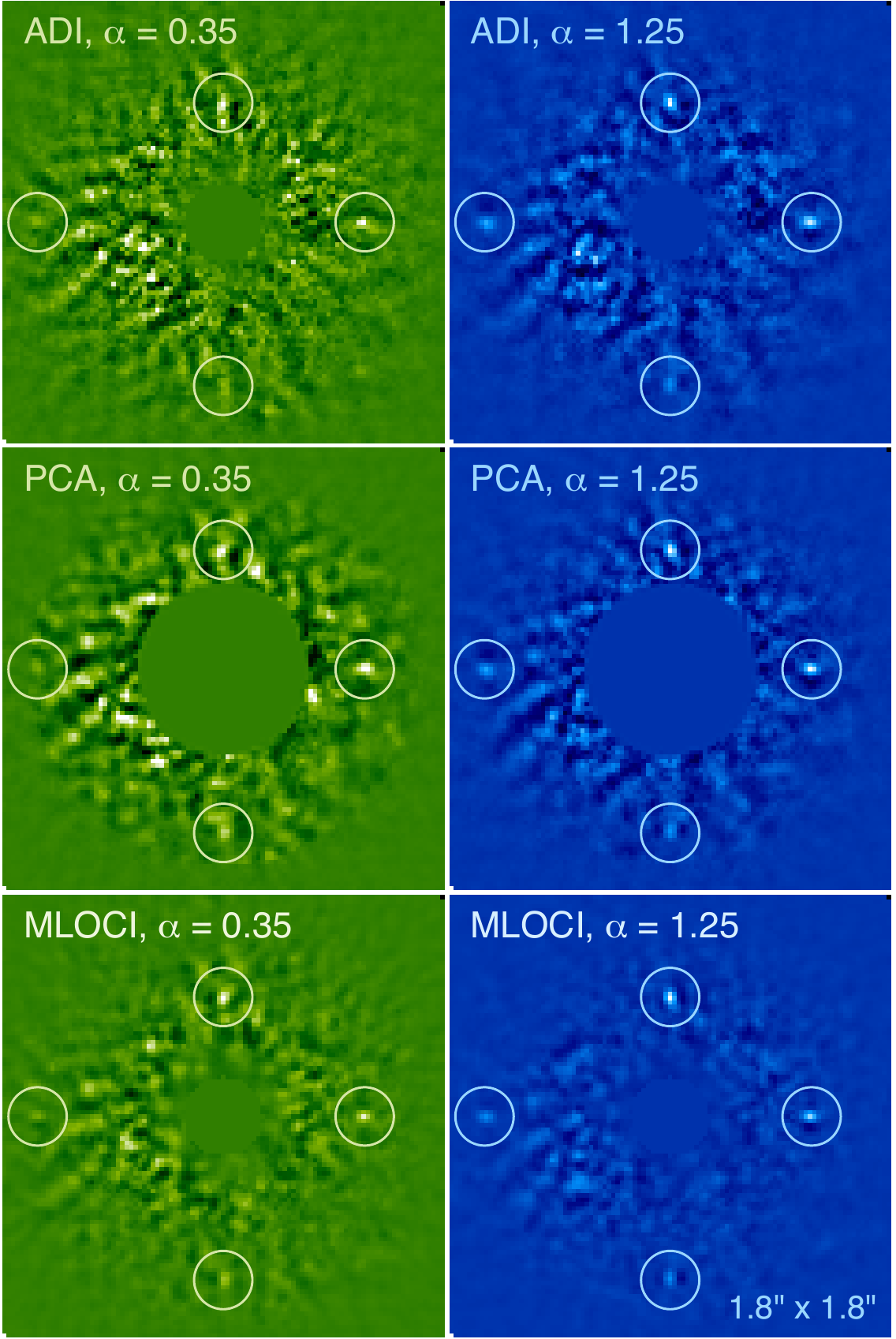}
    }
    \caption{\textcolor{black}{Comparison of MLOCI reductions vs PCA and basic ADI
      reductions of NICI UY Pic data injected with
      simulated companions. The left panels in green show reductions
      for  sky rotation equivalent to  0.35$\times$PSF FWHM ($\alpha=$0.35) of motion
      at 0.$''$5 separation. The blue panels to the right show the
      ones for 1.25$\times$PSF FWHM of motion. The recovered companions can be seen at 
      separations 0.$''$5, 0.$''$6, 0.$''$7 and 0.$''$8 with PA=360,
      270, 180 and 90\dg\ respectively. The images have been normalized to the
      peak flux of the recovered companion at the top position (color
      scale is linear). The SNR improvement for MLOCI over PCA
      were 1.3 to 2.3 for $\alpha=$0.25 and 1.3 to 1.9 for
      $\alpha=$1.25. The SNRs from PCA and basic ADI reductions were
      within 10\% of each other.}}
  \label{fig:compare6}
\end{figure}

\textcolor{black}{
Using the above parameters, we show two MLOCI reductions (for
$\alpha=$0.35 and 1.25) of the UY Pic (K0V, $H$=5.9~mag, age $\sim$ 70~Myr;
\nocite{2006ApJ...643.1160L} L{\'o}pez-Santiago et al.\ 2006) data set along with ADI 
and PCA reductions (see Figure~\ref{fig:compare6}). 
In these reductions, simulated companions were placed at separations 0$''$.5, 0$''$.6, 0$''$.7 and 0$''$.8
and at PAs 360, 270, 180 and 90\dg\ just below the detection limits of
an initial ADI reduction. The SNR improvement factors achieved 
for MLOCI over ADI and PCA range from 1.3 to 2.3. }

In the SNR calculations, for the noise we use the RMS in an annulus of
width 2$\times$FWHM centered on the test locations but exclude circular regions of diameter
1$\times$FWHM around them. 
The signal is just the intensity at the test locations minus the median intensity in the noise region.
To decide whether a source in the reduced images
is a detection based on shape, we calculate the
fractional reduction in the RMS in a box of size 2$\times$FWHM+1
pixels when an optimally scaled Gaussian PSF is subtracted from the source.
Here FWHM is estimated for the source, first.
When the fractional reduction is $>$0.3 and the SNR is 5, we consider
the source a detection, a criteria which recovers a 5$\sigma$ signal
$>$99\% of the time with a negligible false detection rate (FDR;
sometimes termed false discovery rate) in our tests. 
We use this criteria over that in \citet{2013ApJ...779...80W} because it has a clearer interpretation.

\begin{figure}[ht]
  \centerline{
      \includegraphics[width=18cm]{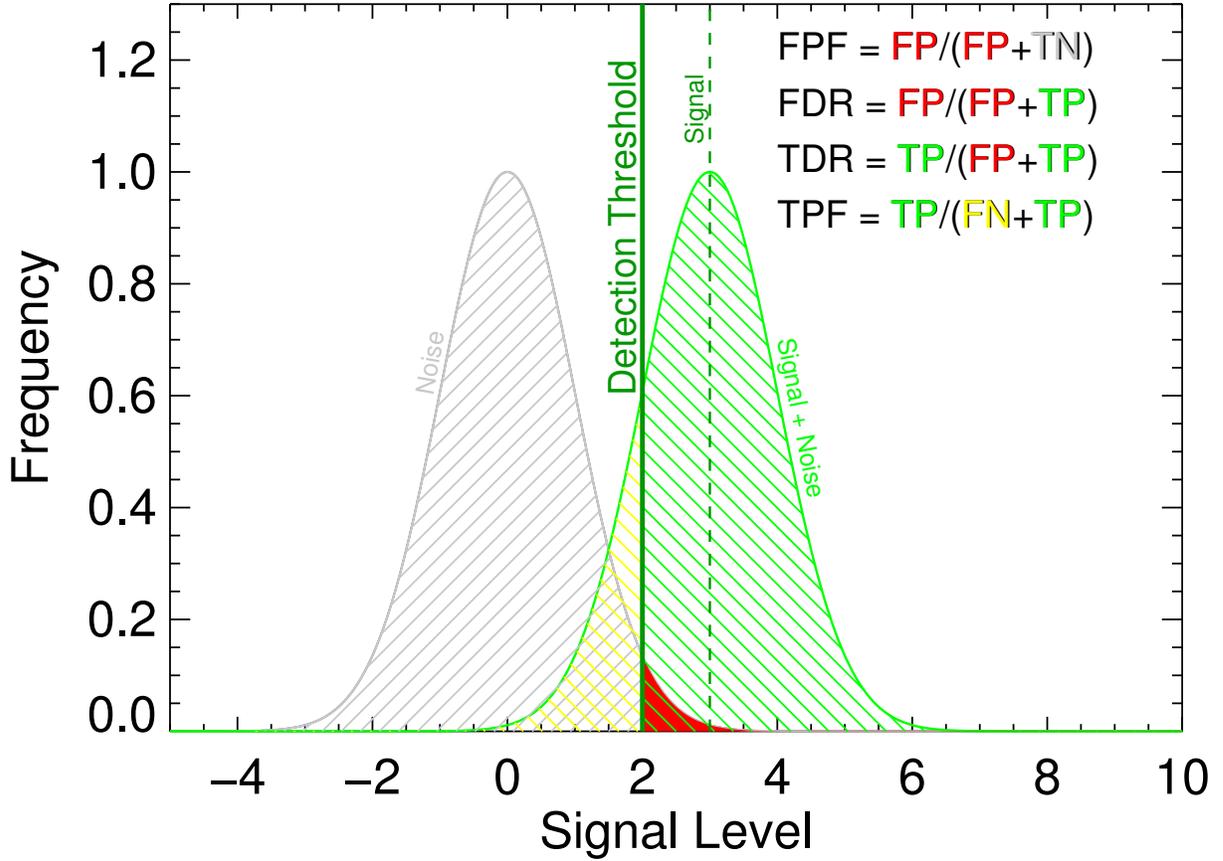}
    }
    \caption{Detection rates for the example case of 3$\sigma$ signals
      and a detection threshold set at 2$\sigma$. Here, we illustrate the relationship between False Positive Fraction (FPF),
      False Detection Rate (FDR), True Detection Rate (TDR) and True
      Positive Rate (TPR or Completeness) and how they depend on 
      True and False Positives (TP shown in green and FP showin in red) and True and False
      Negatives (TN shown in grey and FN shown in yellow). The FDR, which determines the telescope time spent following-up
      bogus detections, should be minimized. This can be quite different in magnitude from the FPR.}
  \label{fig:fdr_tdr}
\end{figure}

\textcolor{black}{
In the example MLOCI reductions of Figure~\ref{fig:compare6}, no other sources besides the ones which were injected into the data were
recovered according to the detections criteria specified above.}
Although MLOCI can improve SNR by a factor $>$2, we need to check that
the FDR is not elevated compared to ADI. The FDR is the fraction of detections that are incorrect, 
i.e., they do not result from real or injected signal. It is important
that this number is low, so that negligible time is wasted in follow-up
observations trying to confirm false detections. The FDR is also
different from FPR, the false positive fraction (see Figure~\ref{fig:fdr_tdr}). 
Next, we discuss FDR for MLOCI reductions.
   
\subsection{ Contrast, SNR,  Completeness and False Detection Rates}
Before presenting an analysis of our detection
statistics, we discuss the relationship between frequently used
terms in the literature. After ADI processing of adaptive optics images, the residual intensity
statistics is nearly Gaussian \citep[][see also for detection statistics at 1--3 $\lambda/D$ from the primary]{2014arXiv1407.2247M}. 
A signal $S$ embedded in Gaussian noise with standard deviation, $\sigma=$1, when detected using 
a signal threshold $T$, will be recovered at a completeness (or true positive rate) of 
$C=G(S-T)=\frac{1}{\sqrt{2\pi}}\int_{T-S}^{\infty} e^{{-x^2}/2} dx$ 
(with inverse function, $G'(C)=S-T$; Equation 1).  
This is because the real signal is redistributed due to the noise in a Gaussian
manner (see Figure~\ref{fig:fdr_tdr}). In other words, the
completeness is the fraction of real (or injected) signal above a level, $S$, that is detected by a method.
The FDR, which we measure as the number of bogus 
detections or the number of detections in the absence of signal, is 
theoretically FPR/(FPR+C), where the FPR is $G(-T)$. 

\textcolor{black}{
Since there is a one-to-one correspondence between SNR improvement and
contrast improvement,
($\Delta$contrast=2.5$\log_{10}\frac{SNR_{new}}{SNR_{old}}$ mag), we only plot
contrast improvements. According to equation 1, an improvement in SNR from 5 to 10 corresponds to
improvement in completeness from 50\% to nearly 100\% (to within
10$^{-6}$), for $T$=5. Thus, completeness improvements quickly become insensitive
to SNR improvements. Nevertheless, since it is interesting to show the
improvement in the fraction of sources recovered, we compare the completeness for PCA and basic ADI at the
threshold where MLOCI yields 50\%. }

\begin{figure}[ht]
  \centerline{
      \includegraphics[width=14cm]{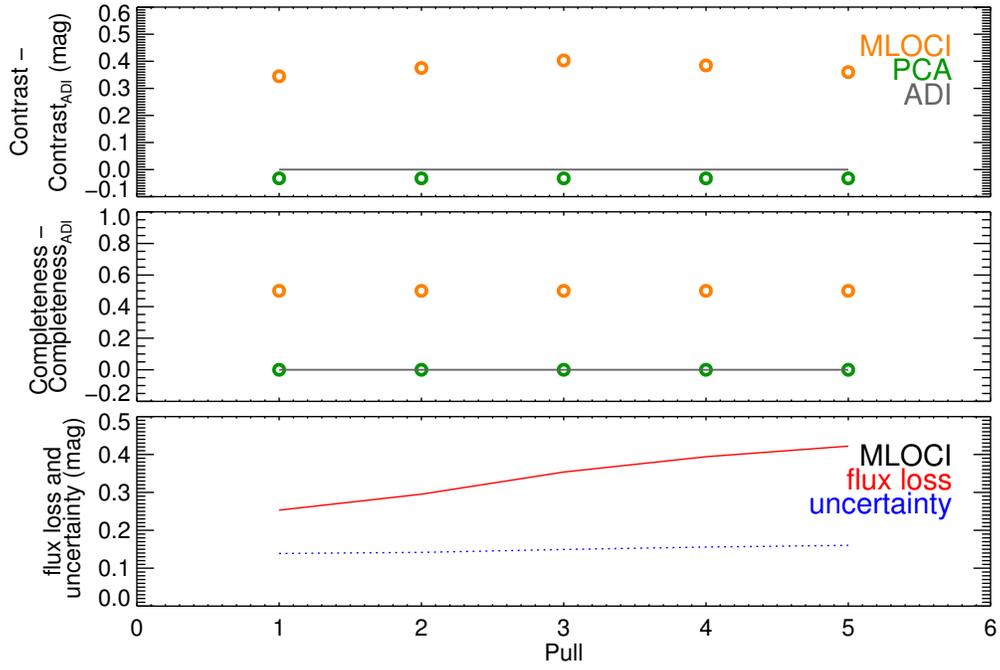}
    }
    \caption{\textcolor{black}{ Comparison of MLOCI to PCA and regular
    ADI for different pull factors (see \S~2). We attempted to recover
    simulated companions with contrasts of \app 9-11 mag at 0$''$.5 separation. We
    compare the 95\%  completeness contrast and completeness improvements for regular ADI, PCA,
    and MLOCI reductions. 
    The subtraction annulus width $W$ is set to 16, while the artificial sky motion $\alpha$
    was set to 2.15$\times$FWHM.}}
  \label{fig:comp_pull}
\end{figure}

\begin{figure}[ht]
  \centerline{
      \includegraphics[width=14cm]{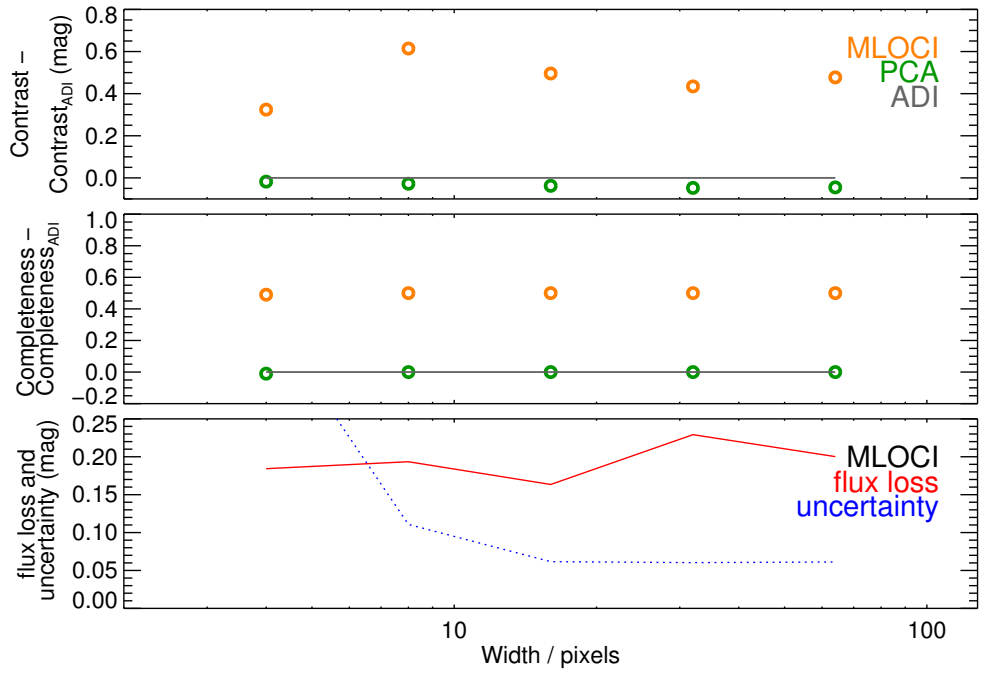}
    }
    \caption{\textcolor{black}{Similar to Figure~\ref{fig:comp_pull}, a comparison of
      MLOCI to PCA and regular ADI for different subtraction region widths and 
    $\alpha$=2.15 and $P$=3}.
  }
  \label{fig:comp_width}
\end{figure}

\begin{figure}[ht]
  \centerline{
      \includegraphics[width=14cm]{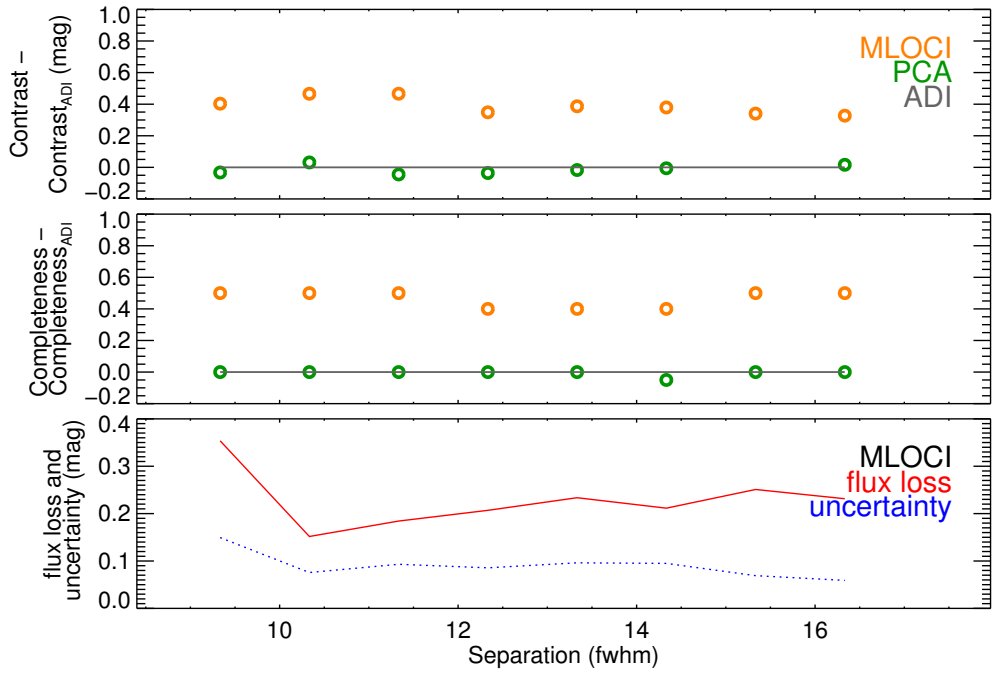}
    }
    \caption{\textcolor{black}{Similar to Figure~\ref{fig:comp_pull}, a comparison of
      MLOCI to PCA and regular
    ADI for different separations and 
    $\alpha$=2.15, $W$=16 and $P$=3.}
  }
  \label{fig:comp_rho}
\end{figure}

\begin{figure}[ht]
  \centerline{
      \includegraphics[width=14cm]{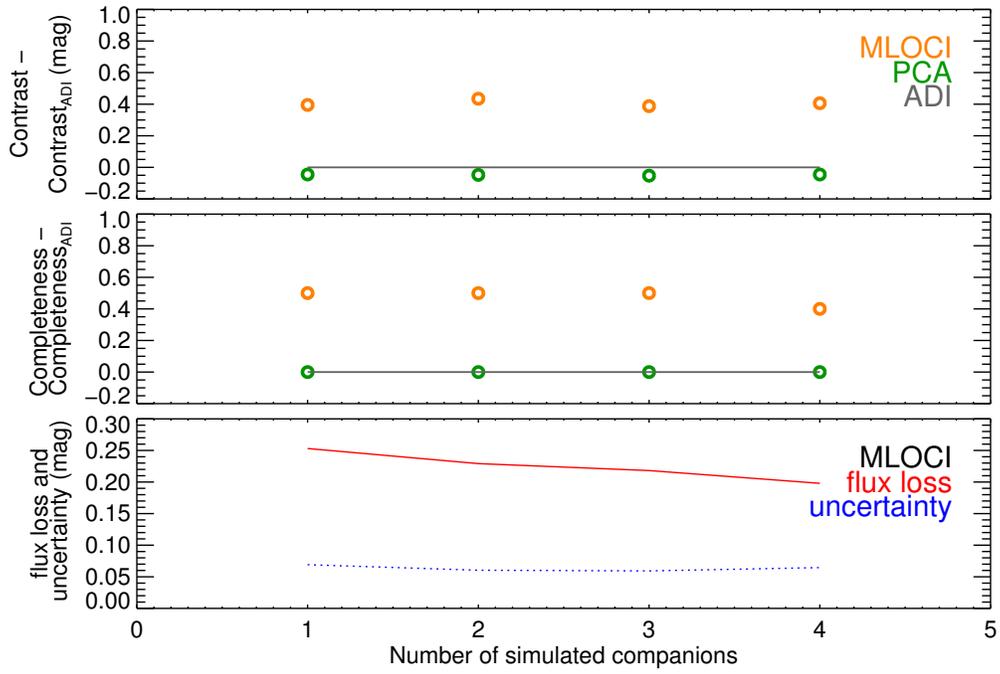}
    }
    \caption{\textcolor{black}{Similar to Figure~\ref{fig:comp_pull}, a comparison of 
    MLOCI to PCA and regular ADI for different number of simulated companions and
    $\alpha$=2.15, $W$=16 and $P$=3. }}
  \label{fig:comp_nn}
\end{figure}

\begin{figure}[ht]
  \centerline{
    \includegraphics[width=14cm]{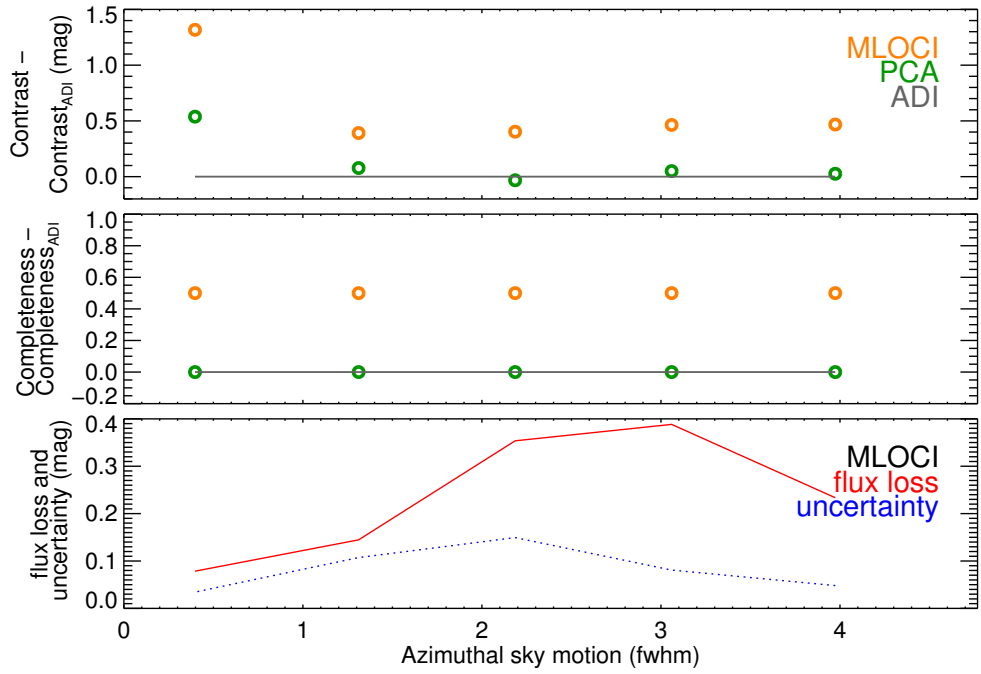}
  }
  \caption{\textcolor{black}{Similar to Figure~\ref{fig:comp_pull}, a comparison of
    MLOCI to PCA and regular
    ADI for different amounts of sky rotation and $P$=3 and
    $W$=16.}
  }
  \label{fig:comp_paf}
\end{figure}

\textcolor{black}{
In Figures~\ref{fig:comp_pull}--\ref{fig:comp_paf}, we compare average detection statistics from our 10 NICI datasets for
basic ADI, PCA  and MLOCI  for different reduction parameters.
In Figure~\ref{fig:comp_pull}, we see that the performance of MLOCI is
not very sentitive to $P$, the relative intensity of the negative sources, generally yielding contrast
improvements of 0.3-0.4~mag. The plots are made for $\alpha$=2.15 and
optimum parameter settings $W$=16 and $N$=2.  
However, for very large $P$ (\eg\ 10--100) we have found that the
performance of MLOCI does deteriorate. Moreover, the counterpart PSFs begin to look similar to each other, instead
of the science images that they are being matched to. This is because 
LOCI strongly biases the linear combination to match the negative sources in
the counterpart images. The completeness improvement for MLOCI at the
chosen threshold is usually 50\%. The ADI and PCA performances are nearly equal
to each other. }

\textcolor{black}{
From our reduction experiments, we can estimate the accuracy of the 
photometric measurements allowed by MLOCI. The systematic uncertainty
of the photometry can be estimated by the change in 
the contrast before and after recovery, between the companion in the
test sector to the two point sources in the
reference sector. Since the reduction
is done 10 times with the reference companions in different positions,
we can also estimate the random uncertainty by calculating the
standard deviation of the contrast after recovery between the two sets of companions. When an actual
companion is detected in a real dataset, we can estimate photometric
uncertainties and systematic errors by injecting companions at other PAs at the same
separation and performing MLOCI reductions to recover them.
In the bottom panel of Figure~\ref{fig:comp_pull}, we see that while
the photometric uncertainty is insentitive to $P$ (variation of $\sim$13\% in these
cases), the systematic error increases with increasing $P$.
 }

\textcolor{black}{
In Figures~\ref{fig:comp_pull}--\ref{fig:comp_rho}, we see that the 
contrast (and completeness) improvements are not very sensitive to the width of the 
reference sector or the number of companions used as reference, or the
separation of the recovered companions from the primary. However,
Figure~\ref{fig:comp_paf} shows that for very small sky rotation
($\alpha=$0.35) the contrast gain due to MLOCI is much larger
($\sim$1.4~mag). Here, PCA also shows a 0.5~mag gain over ADI. 
The variation of photometric uncertainties over these experiments is small.
}

\begin{figure}[ht]
  \centerline{
    \includegraphics[width=18cm]{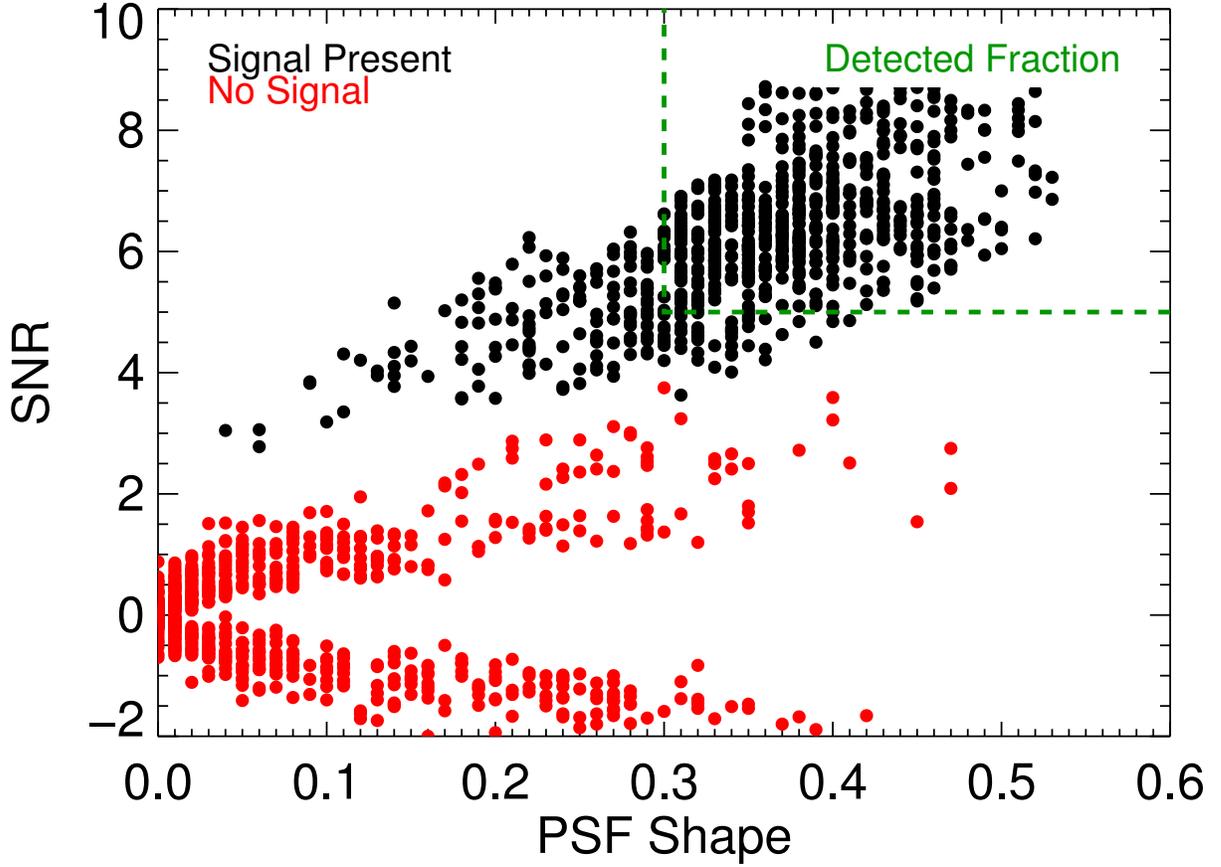}
  }
  \caption{ \textcolor{black}{The black dots show the measured signal to noise and PSF
    shape (same as fractional reduction in \S4.1) of point sources injected at tests spots
    after 800 MLOCI reductions. The green dashed lines show the
    detection criteria defined in \S4. The red dots show 
    measurements for another 800 reductions where no signal was injected. No false positives
    were found; the red dots lie outside the green box and reasonably
    far from the detection limits. Moreover, the standard deviations of
    the black and red SNR distributions are 1.2 and 1 respectively.}
  }
  \label{fig:false_pos}
\end{figure}

\textcolor{black}{
In Figure~\ref{fig:false_pos}, we show attempts at measuring the source
properties at the test locations after 800 MLOCI reductions of the UY
Pic data set when no actual signal has been injected. According to the previously defined 
detection criteria in \S4, none of the measurements satisfied the
detection crtieria and thus no false positives were detected. 
Moreover, the standard deviation of the SNR distribution  was 1.0,
showing that the false positive rate is not elevated above that expected.
A higher false positive rate would indicate that while the reduction
method may produce detections with seemingly high SNR, the noise is 
in fact not being calculated correctly.   
}

\textcolor{black}{
We also performed non-blind reduction experiments where the test companion and the
reference companions are treated the same by MLOCI, using negative
sources in all locations when creating the counterpart PSFs and/or
positive sources when combining all the images. In these cases, the
apparent contrast improvement was on average 0.75~mag. However, using the best 
algorithm parameters we were only able to reduce the false detection rate to 1\%. 
Poor parameters could yield false detection rates $>$ 20\%. Thus, MLOCI must be kept blind about any suspected companion,
except when informed by another blind process, which in our case is 
a basic ADI reduction. By the same logic, other versions of LOCI and PCA can also accidentally create false detections,
since no restrictions on the allowed linear combinations are generally
imposed when constructing reference images. This is an especially
unpleasant possibility when trying to re-detect a companion marginally
detected in an earlier epoch. We therefore strongly recommend that 
FDRs be provided for any reduction method.
}
  
\subsection{GPI Early Science Data}

\begin{figure}[ht]
  \begin{center}
    \includegraphics[height=18cm]{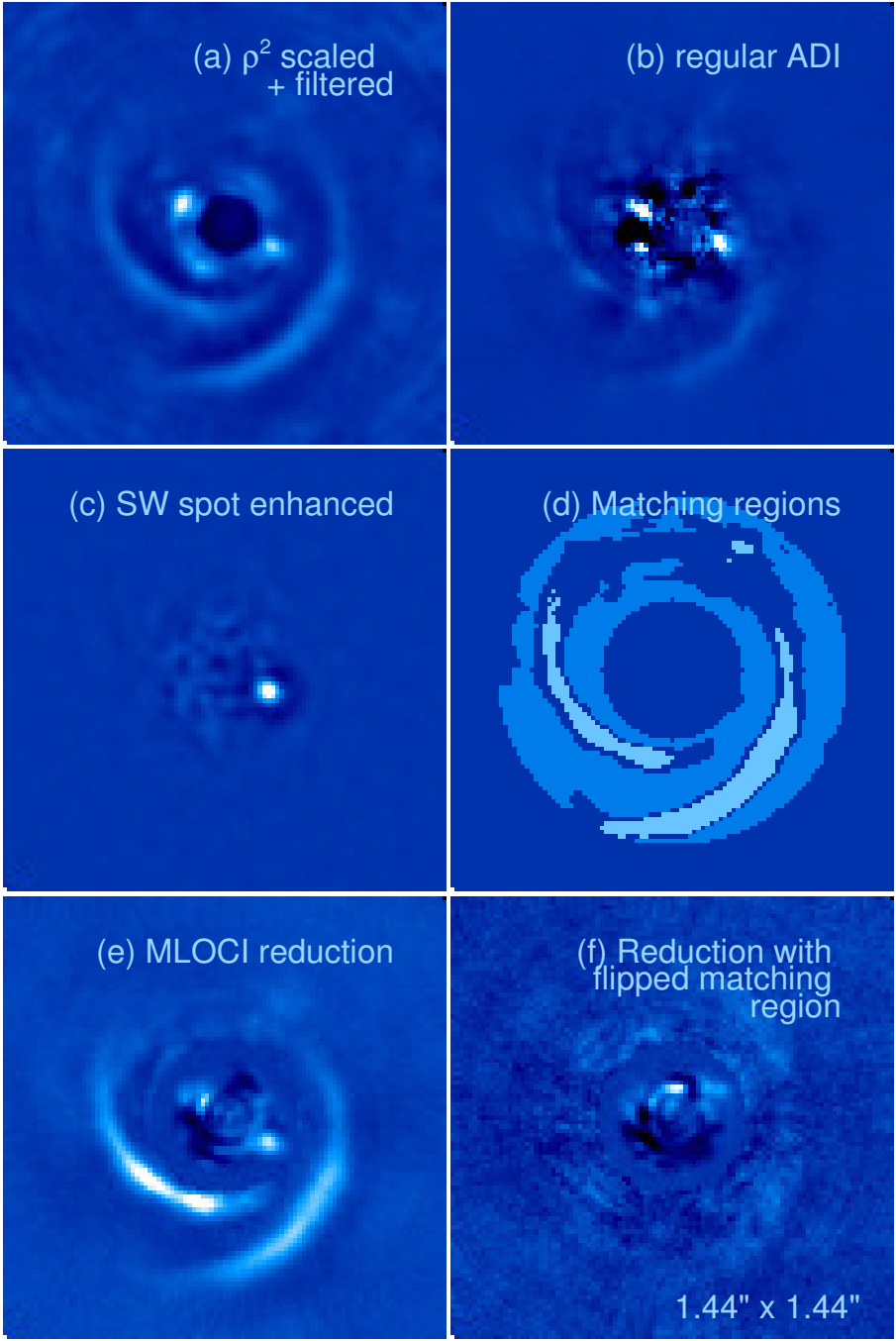}
    \end{center}
    \caption{\textcolor{black}{All images are normalized by
        \textcolor{black}{the RMS over the full images}. They 
        are oriented are North up and East to the left.
        {\bf(a)} Simple stack of the \prim\ GPI $J$-band dataset 
        scaled by $\rho^2$ and spatially filtered \textcolor{black}{to isolate 
          features between 2--8 pixels in size}. 
        {\bf(b)} Regular ADI reduction  
        {\bf(c)} Channel~35 ($\lambda$=1.337\mic ) reduced with MLOCI X, enhancing
        only the SW spot.  
        {\bf(d)} MLOCI matching regions: Signal is preserved in the
        lightest regions. Noise is suppressed in medium blue
        regions. Darkest regions are not matched.
        {\bf(e)} MLOCI reduction matched to regions in {\bf d}. Region
        within 240~mas of center divided by 5, to improve stretch.  
        {\bf(f)} MLOCI reduction matched to map in {\bf d} but flipped
        vertically. This demonstrates that it is difficult to create
        false disks by MLOCI.}
    }   
  \label{fig:gpi_reducs}
\end{figure}

Here, we describe the results of MLOCI and ADI processing of the GPI data on \prim . 
In Figure~\ref{fig:gpi_reducs}a, we show a de-rotated
median combination of all the
channels without PSF subtraction, but scaled in intensity by
\ro$^2$, and then bandpass filtered to isolate 
spatial features between 2--8 pixels. We easily see the spiral arms, discovered
in \citet{2012ApJ...748L..22M},  extending out to 0.9$''$, and the cavity
inside 0.2$''$. However, we do not see their inner ring where
the spiral arms begin. We do see a bright rim at 0.15$''$ separation around the coronagraphic mask. 
Since, there is presumably no self-subtraction of the disk (no ADI) in this image,
we use this image to identify reliable signal and background regions
for MLOCI  (Figure~\ref{fig:gpi_reducs}d).

In the regular ADI reduction (Figure~\ref{fig:gpi_reducs}b) we find some features at 110--170~mas
separations with SNR=3--5 ($\Delta J$=8-9.5 mag), particularly two radially extended spots to the NE and
SW. As discussed in Menard et al.\ (2015, in preparation), while these could prove to
be accretion streams from the disk onto the star, this is currently uncertain because of the small
amount of sky motion (1.7--2$\times$FWHM) at the relevant separation and similar PSF features
in the unreduced images near mask (see Figure~\ref{fig:gpi_reducs}a).  Moreover, GPI images
of HR~4796~A \citep{2014arXiv1407.2495P} also show similar residuals near the mask.

We find that particular caution needs to be exercised with MLOCI when there is
signal in the FoV which is extended azimuthally, as in the case of
\prim . When we try to target each spot separately with non-blind MLOCI, we obtain isolated point
sources, admittedly with large FWHM, while the rest of the disk
disappears (see Figures~\ref{fig:gpi_reducs}c). This demonstrates the
ability of non-blind algorithms which
allow free linear combinations of science images to produce misleading
features, especially in the case where disk emission may
exist. In the absence of disk emission, we have found that the most
conservative non-blind methods create less false positives but still
at elevated levels compared to ADI.

\textcolor{black}{
However, the MLOCI algorithm for extended-sources does have the special ability 
to improve the SNR of any feature suspected 
in a dataset, be it mutiple companions or an entire disk. 
To demonstrate this potential of the algorithm, we use MLOCI ($P$=3; 
extended-source reduction as described in \S2 \& \S3) to optimize
the SNR according to regions selected in Figure~\ref{fig:gpi_reducs}d.
Indeed, the spiral and the suspected streamers are much enhanced in
the MLOCI reduction (see Figure~\ref{fig:gpi_reducs}e). When we flip the matching regions vertically 
so that they do not match with the real disk emission, MLOCI yields
no disk (Figure~\ref{fig:gpi_reducs}f). Thus MLOCI is not likely to
yield false disk detections.}

\section{CONCLUSIONS}
We have presented a version of the LOCI algorithm that significantly improves the SNR of
both point and extended sources. This algorithm, called Matched LOCI
or MLOCI, can be applied to both ADI and IFS
data. The only difference being that, in the PSF construction
step of image processing, the images have to be pupil aligned in the
former and both  pupil and spectrally aligned in the latter. 
The contrast gain over both regular ADI and PCA in NICI
data was 0.3-1.4 mag. This is especially interesting, since it was shown that LOCI itself
provides no gain over regular ADI processing for NICI data \citep{2013ApJ...779...80W}. 
While we found that LOCI could reduce the speckle 
noise in the final reduced image, it also reduced the signal of test companions.

We have shown how MLOCI can be used in the detection of disks
using GPI data, recovering new streamer-like features 
which need new observations before they can be confirmed. 
Thus MLOCI can be a powerful tool for the detection of companions 
and disks in the current direct imaging campaigns with GPI and SPHERE.

While, for blind MLOCI the false detection rate is low,
we found that for non-blind MLOCI for both NICI and GPI data, the false detection
rate (FDR) can in some cases be much higher. The FDR is even
more for when there is azimuthally distributed diffuse light in the FoV. 
Thus, in non-blind variants of LOCI and PCA algorithms high
FDRs may also arise. Therefore, we strongly recommend that FDRs be measured whenever these methods are used.

\begin{acknowledgements}  
Our research has employed the 2MASS data products; NASA's Astrophysical
Data System; the SIMBAD database operated at CDS, Strasbourg, France.
L.A.C. was supported by ALMA-CONICYT grant number 31120009 and CONICYT-FONDECYT grant number 1140109. 
L.A.C., M.R.S., H.C., and S.C. acknowledge support from the Millennium Science Initiative (Chilean Ministry of Economy),
through grant “Nucleus P10-022-F”.   
\end{acknowledgements} 

{\it Facilities:} Gemini-South (GPI,  NICI).

\vfill
\eject

\bibliographystyle{apj}
\bibliography{zrefs}

\vfill
\eject


\end{document}